\documentclass[floats, prd, eqnum, showpacs, nofootinbib, 
twocolumn, 
eqsecnum]{revtex4-1}

\usepackage{color,graphicx}
\usepackage{amsfonts}
\usepackage{amssymb}
\begin{document}

\title{Circular light orbits of a general, static, and spherical symmetrical wormhole with $Z_2$ symmetry}
\author{Naoki Tsukamoto${}^{1}$}\email{tsukamoto@rikkyo.ac.jp}

\affiliation{
${}^{1}$Department of Physics, Faculty of Science, Tokyo University of Science, 1-3, Kagurazaka, Shinjuku-ku, Tokyo 162-8601, Japan 
}

\begin{abstract}
Recently, the ring images or the shadow images of the centers of the galaxy M87 and the Milky way have been reported by Event Horizon Telescope Collaboration.
It is believed that the ring images imply that the central objects form unstable light circular orbits.
Some of wormholes with $Z_2$ symmetry against a throat are wrongly excluded from the candidates at the centers of M87 and the Milky way due to the overlooking the unstable light circular orbits on the throat.
A general asymptotically-flat, static, and spherical symmetrical wormhole without a thin shell has at least one unstable circular light orbit at the throat or elsewhere. 
If the wormhole has $Z_2$ symmetry against the throat, it has the unstable circular light orbits on the throat or it has stable circular light orbits on the throat and unstable ones near the throat. We need to analyze the throat carefully to make sure we do not unfairly rule out the $Z_2$-symmetrical wormholes.
In this study, we categorize the numbers of the circular light orbits of the $Z_2$-symmetrical wormhole and their stability from the derivatives of an effective potential at the throat 
and we investigate the circular light orbits around a Simpson-Visser black-bounce spacetime, a Damour-Solodukhin wormhole spacetime, a Reissner-Nordstr\"{o}m black-hole-like wormhole spacetime or a charged Damour-Solodukhin wormhole spacetime as examples. 
We give complete treatments including degenerated circular light orbits made from more than one stable and unstable circular light orbits on and off the throat.    
\end{abstract}

\maketitle

\section{Introduction}
In general relativity, a static and spherically symmetric compact object has a photon (antiphoton) sphere which is a sphere formed by unstable (stable) circular light orbits~\cite{Claudel:2000yi,Perlick_2004_Living_Rev}   
and rays can be deflected strongly near the photon sphere~\cite{Hagihara_1931,Darwin_1959,Atkinson_1965,Luminet_1979,Ohanian_1987,Nemiroff_1993,Frittelli_Kling_Newman_2000,Virbhadra_Ellis_2000,Bozza_Capozziello_Iovane_Scarpetta_2001,Bozza:2002zj,Eiroa:2002mk,Perlick:2003vg,Bozza_2010,Tsukamoto:2016jzh,Shaikh:2019itn}.
The existence of the circular light orbits has a strong influence on the image of a collapsing star~\cite{Ames_1968,Synge:1966okc,Yoshino:2019qsh},
the high-frequency behavior of the photon absorption cross section~\cite{Sanchez:1977si,Decanini:2010fz},
the high-frequency spectrum of quasinormal modes of compact objects~\cite{Press:1971wr,Goebel_1972,Raffaelli:2014ola},
centrifugal force and gyroscopic precession~\cite{Abramowicz_Prasanna_1990,Abramowicz:1990cb,Allen:1990ci,Hasse_Perlick_2002},
Bondi's sonic horizon of a fluid~\cite{Mach:2013gia,Chaverra:2015bya,Cvetic:2016bxi,Koga:2016jjq,Koga:2018ybs,Koga:2019teu},
and an ergoregion~\cite{Ghosh:2021txu}. 
It is concerned that stable circular light orbits might cause instability of compact objects with a trivial topology~\cite{Keir:2014oka,Cardoso:2014sna,Cunha:2017qtt,Cunha:2017eoe,Cunha:2022gde,Zhong:2022jke}.

Out of theoretical interests, theorems on the circular light orbits such as  
the upper and lower bounds of radii of circular light orbits around a black hole~\cite{Hod:2017xkz,Hod:2020pim,Hod:2023jmx} and 
the numbers of circular light orbits surround compact objects without an event horizon~\cite{Hod:2017zpi} were investigated. 
The theorems can be applied for the compact objects with a trivial topology in general relativity under energy conditions.
A bound on the size of the circular light orbit around a black hole in an Einstein-Gauss-Bonnet theory under the energy conditions was also shown in~\cite{Ghosh:2023kge}.
Recently, Kudo and Asada have proved that a spacetime cannot be asymptotically flat if its outermost circular light orbit is stable~\cite{Kudo:2022ewn}.
We note that Kudo and Asada's proof relies on only the assumptions on a metric and it depends on neither the energy conditions, gravitational theories, nor the topology of the spacetime. 

For a long time, such a strong gravitational field predicted by general relativity, which was unlikely to be observed, had been not considered seriously in astrophysics.   
Recently, however, phenomena in a strong gravitational field 
has been considered eagerly since the direct detection of gravitational waves from binary black holes have been reported by 
LIGO Scientific Collaboration and Virgo Collaboration~\cite{Abbott:2016blz} and    
the black hole shadows in the centers of a galaxy M87 and the Milky way have been reported by Event Horizon Telescope Collaboration~\cite{Akiyama:2019cqa,EventHorizonTelescope:2022wkp}. 

General relativity permits wormhole spacetimes with a nontrivial topology 
since the Einstein equations are local-field equations which do not determine the topology of the spacetime.
The traversable wormholes have a throat which connects two regions of one or two universes~\cite{Visser_1995,Morris:1988cz}.
In Ref.~\cite{Perlick_2004_Living_Rev}, Perlick pointed out that every asymptotically-flat, static, and spherical symmetrical, traversable wormhole without a thin shell has at least one photon sphere at some radius. 
It comes from assumptions that a metric is of class $C^1$ everywhere and that the spacetime is the asymptotically flat and the latter can be relaxed~\cite{Perlick}. 

Recently,
Shaikh \textit{et al.}~\cite{Shaikh:2018oul} have proved that 
a general, static, and spherically symmetric wormhole with $Z_2$ symmetry against the throat
has unstable or stable circular light orbits on the throat and they have investigated the condition 
when the wormhole has three photon spheres, where one of them is on the throat and the others are off the throat, 
and two antiphoton spheres off the throat.   
Bronnikov and Baleevskikh~\cite{Bronnikov:2018nub} 
also have given an alternative proof on circular light orbits on the throat of the $Z_2$-symmetrical wormhole  
and they have investigated deflection angle of a light scatted by an asymmetric wormhole.
They are interested in the $Z_2$-symmetrical wormhole with 
the photon sphere on the throat~\cite{Shaikh:2018oul,Bronnikov:2018nub} while 
a few $Z_2$-symmetrical wormholes with the antiphoton sphere 
on the throat have been studied also in Refs.~\cite{Shaikh:2019jfr,Tsukamoto:2021apr,Tsukamoto:2021caq}. 

The photon sphere and the antiphoton sphere on the throat of the $Z_2$-symmetrical wormhole is often overlooked. 
In Ref.~\cite{Dey:2008kn}, 
Dey and Sen claimed that
 a $Z_2$-symmetrical Ellis-Bronnikov wormhole~\cite{Ellis:1973yv,Bronnikov:1973fh}  
with vanishing Arnowitt-Deser-Misner(ADM) masses has no photon sphere on its throat
but it contradicts papers by several authors~\cite{Ellis:1973yv,Chetouani_Clement_1984,Perlick:2003vg,Perlick_2004_Living_Rev,Muller:2004dq,Nandi:2006ds,Muller:2008zza,Bhattacharya:2010zzb,Gibbons:2011rh,Nakajima:2012pu,Tsukamoto:2012xs,Ohgami:2015nra,Tsukamoto:2016zdu,Tsukamoto:2016qro,Tsukamoto:2017edq,Bugaev:2023mlc}.
Simpson and Visser suggested a black-bounce spacetime~\cite{Simpson:2018tsi}
has an antiphoton sphere, a degenerated photon sphere, or a photon sphere on the throat as shown in Ref.~\cite{Tsukamoto:2020bjm}
while the antiphoton sphere and the photon sphere on the throat were overlooked in Refs.~\cite{Simpson:2018tsi,Nascimento:2020ime,Guerrero:2021ues,Bronnikov:2021liv,Combi:2024ehi}.
We call a degenerated photon sphere (degenerated antiphoton sphere) a sphere filled in marginally unstable (stable) circular light orbits 
because they are made from more than one photon spheres and antiphoton shpheres.        
A Damour-Solodukhin wormhole~\cite{Damour:2007ap} also have 
the antiphoton sphere, the photon sphere, or the degenerated photon sphere on the throat~\cite{Tsukamoto:2020uay,Tsukamoto:2023rzd}.
However, the (anti)photon sphere on the throat is often neglected~\cite{Nandi:2018mzm,Ovgun:2018fnk,Bhattacharya:2018leh,Ovgun:2018swe,Ovgun:2018oxk,Vagnozzi:2022moj}.

The wormhole spacetimes are defined by the existence of the throat.
Thus, the overlooking the circular light orbit on the throat can cause the serious problems on investigation of the wormholes.
In a dozen papers~\cite{Simpson:2018tsi,Nascimento:2020ime,Guerrero:2021ues,Bronnikov:2021liv,Combi:2024ehi,Nandi:2018mzm,Ovgun:2018fnk,Bhattacharya:2018leh,Ovgun:2018swe,Ovgun:2018oxk,Vagnozzi:2022moj,Dey:2008kn}, 
the photon sphere on the throat was neglected. 
Before the shadow observations, these overlooks were not significant problems.  
Recently, however, some of the wormholes are wrongly excluded from the candidates at the centers of the galaxy M87 and the Milky way due to the overlooking the photon spheres since
the ring images or the shadows of the centers of the M87 and the Milky way imply that they have the photon spheres.
Our purpose in this paper is to draw researchers' attention to the issue.    

In this work, by combining Kudo and Asada's theorem~\cite{Kudo:2022ewn} 
and the theorem on the $Z_2$-symmetrical wormhole in Ref.~\cite{Shaikh:2018oul,Bronnikov:2018nub},
we show a fact that a general asymptotically-flat, static, and spherically symmetrical wormhole 
with $Z_2$ symmetry against the throat has the unstable circular light orbits on the throat or it has stable circular light orbits on the
throat and unstable ones near the throat.
\footnote{From the symmetry of static, spherically symmetrical, and $Z_2$-symmetrical wormholes, 
we can easily show that they have the circular light orbits on the throat. 
But the $Z_2$-symmetrical wormholes with stable light circular orbits on the throat and without any unstable light orbits might be permitted 
if the asymptotic flatness is not imposed.}
We would get it also by applying Perlick's theorem~\cite{Perlick_2004_Living_Rev} for the $Z_2$-symmetrical wormhole.
It helps to prevent the overlook the photon spheres and the antiphoton spheres of the wormhole. 
We confirm the theorems in the Simpson-Visser black-bounce spacetime~\cite{Simpson:2018tsi}, the Damour-Solodukhin wormhole spacetime~\cite{Damour:2007ap}, 
and a Reissner-Nordstr\"{o}m black-hole-like wormhole spacetime~\cite{Lemos:2008cv,Tsukamoto:2019ihj}.

This paper is organized as follows.  
In Sec.~II, we consider the photon spheres and antiphoton spheres
in a general asymptotically-flat, static, and spherically symmetrical wormhole spacetime
with $Z_2$ symmetry against the throat.
We consider the Simpson-Visser black-bounce spacetime~\cite{Simpson:2018tsi}, the Damour-Solodukhin wormhole spacetime~\cite{Damour:2007ap}, 
and the Reissner-Nordstr\"{o}m black-hole-like wormhole spacetime~\cite{Lemos:2008cv,Tsukamoto:2019ihj} as examples in Sec.~III, 
and we summarize and discuss in Sec.~IV. 
In this study, we use units in which a light speed and Newton's constant are unity.

\section{circular light orbits on a throat}
We consider a general, static, and spherical symmetrical wormhole spacetime with $Z_2$ symmetry against a throat. 
Its metric is given by, in coordinates~$(t, x, \theta, \varphi)$,
\begin{eqnarray}
\mathrm{d}s^2=-A(x) \mathrm{d}t^2+\frac{\mathrm{d}x^2}{A(x)} +r^2(x)\left( \mathrm{d}\theta^2+ \sin^2 \theta \mathrm{d}\varphi^2 \right), \nonumber\\
\end{eqnarray}
where the radial coordinate $x$ is defined for $-\infty<x<\infty$.
We assume that $A(x)$ and $r(x)$ are positive, continuous, and finite for $-\infty<x<\infty$.
The spacetime has time-translational and axial Killing vectors 
$t^\mu \partial_\mu=\partial_t$ and $\varphi^\mu \partial_\mu = \partial_\varphi$ due to its stationality and axisymmetry, respectively. 
We set $\theta=\pi/2$ without loss of generality because of spherically symmetry.
We assume that $r(x)$ has a positive minimal value at $x=0$, i.e., there is the throat at $x=x_\mathrm{th}=0$.
We also assume that the wormhole has a $Z_2$ symmetry against the throat.  
From the assumptions, we obtain $A^{\prime}(x_\mathrm{th})=r^{\prime}(x_\mathrm{th})=0$ and $r^{\prime \prime}(x_\mathrm{th})\geq 0$, 
where the prime denotes a differentiation with respect to $x$.

The trajectory of a ray is expressed by 
\begin{eqnarray}\label{eq:trajectory}
-A(x) \dot{t}^2+\frac{\dot{x}^2}{A(x)} +r^2(x) \dot{\varphi}^2=0,
\end{eqnarray}
where the dot denotes a differentiation with respect to an affine parameter along the trajectory.
By using the impact parameter $b\equiv E/L$ of the ray, where $E\equiv -g_{\mu \nu}t^\mu \dot{x}^\nu$
 and $L\equiv g_{\mu \nu} \varphi^{\mu} \dot{x}^\nu$ are conserved energy and angular momentum of the ray, respectively,
Eq.~(\ref{eq:trajectory}) can be rewritten in 
\begin{eqnarray}\label{eq:trajectory2}
\dot{x}^2+V_\mathrm{eff}(x,b)=0,
\end{eqnarray}
where $V_\mathrm{eff}(x,b)$ is the effective potential of the ray given by
\begin{eqnarray}
V_\mathrm{eff}(x,b)\equiv E^2 A(x) \left( \frac{b^2}{r^2(x)} -\frac{1}{A(x)} \right).
\end{eqnarray}
The first and second derivatives of the effective potential are obtained as 
\begin{eqnarray}
V_\mathrm{eff}^{\prime}(x,b)= E^2 b^2 \left( \frac{A^\prime r -2Ar^\prime}{r^3} \right)
\end{eqnarray}
and
\begin{eqnarray}
V_\mathrm{eff}^{\prime \prime}(x,b)=  E^2 b^2 \left( \frac{A^{\prime \prime} r^2 -4 A^\prime r r^\prime +6A r^{\prime 2} -2Ar r^{\prime \prime} }{r^4} \right). \nonumber\\
\end{eqnarray}

If the spacetime permits
\begin{eqnarray}
A^\prime (x_\mathrm{c}) r(x_\mathrm{c}) -2A(x_\mathrm{c})r^\prime(x_\mathrm{c})=0,
\end{eqnarray}
where $x_\mathrm{c} \neq 0$,
and $V_\mathrm{eff}^\prime(x_\mathrm{c},b)=0$ holds,
the wormhole has circular light orbits at $x=x_\mathrm{c}$ where is off the throat.
The ray on the circular orbit has a critical impact parameter $b=b_\mathrm{c}$, where $b_\mathrm{c}$ is given by 
\begin{eqnarray}
b_\mathrm{c} \equiv \pm \sqrt{\frac{r^2 (x_\mathrm{c})}{A(x_\mathrm{c})}}, 
\end{eqnarray}
so that $V_\mathrm{eff}(x_\mathrm{c},b_\mathrm{c})$ vanishes.
The circular light orbit is unstable (stable) if $V_\mathrm{eff}^{\prime \prime}(x_\mathrm{c}, b_\mathrm{c})$ is negative (positive). 
A sphere formed by the unstable (stable) circular light orbits is called photon sphere (antiphoton sphere).
Due to the $Z_2$ symmetry against the throat, 
the wormhole has circular light orbits at $x=\pm x_\mathrm{c}$ off the throat in the both regions against the tharot.

We find $V_\mathrm{eff}^{\prime}(x_\mathrm{th},b)=0$ 
and $V_\mathrm{eff}(x_\mathrm{th},b_\mathrm{th})=0$, where $b_\mathrm{th}$ is a critical impact parameter
\begin{eqnarray}
b_\mathrm{th} = \pm \sqrt{\frac{r^2 (x_\mathrm{th})}{A(x_\mathrm{th})}}, 
\end{eqnarray}
on the throat $x=x_\mathrm{th}=0$.
Thus, the rays make the photon sphere (antiphoton sphere) on the throat
if $V_\mathrm{eff}^{\prime \prime}(x_\mathrm{th}, b_\mathrm{th})$ is negative (positive).
Here, $V_\mathrm{eff}^{\prime \prime}(x_\mathrm{th},b)$ is obtained by 
\begin{eqnarray}
V_\mathrm{eff}^{\prime \prime}(x_\mathrm{th},b)=  E^2 b^2 \left( \frac{A^{\prime \prime}(x_\mathrm{th})  r(x_\mathrm{th}) -2A(x_\mathrm{th}) r^{\prime \prime}(x_\mathrm{th}) }{r^3(x_\mathrm{th})} \right). \nonumber\\
\end{eqnarray} 

If the wormhole spacetime is asymptotically flat at both regions against the throat, 
\begin{eqnarray}
&&A(\pm \infty)= 1 + O\left( x^{-1} \right) = 1 + O\left( r^{-1} \right), \\
&&A^\prime (\pm \infty)= O\left( x^{-2} \right) = O\left( r^{-2} \right), \\
&&r(\pm \infty)= \pm x + O\left( x^{0} \right) 
\end{eqnarray} 
should hold and, hence, we obtain 
\begin{eqnarray}
V_\mathrm{eff}^{\prime}(\pm \infty,b)= \pm  \frac{2 E^2 b^2}{r^3 (\pm \infty)} .
\end{eqnarray}
From $V_\mathrm{eff}^{\prime}(-\infty,b)>0$, $V_\mathrm{eff}^{\prime}(x_\mathrm{th},b)=0$, and $V_\mathrm{eff}^{\prime}(\infty,b)<0$, and an intermediate value theorem,
we obtain the following theorem on the asymptotically-flat wormhole;
\begin{description} 
 \item[(a)]For $V_\mathrm{eff}^{\prime \prime}(x_\mathrm{th},b)<0$, the numbers of the photon spheres and antiphoton spheres are $2n-1$ and $2n-2$, respectively, where $n$ is a positive integer. 
One of the photon spheres is on the throat.
 \item[(b)]For $V_\mathrm{eff}^{\prime \prime}(x_\mathrm{th},b)>0$, the numbers of the photon spheres and antiphoton spheres are $2n$ and $2n-1$, respectively.
One of the antiphoton spheres is on the throat.
 \item[(c)]The photon sphere and the antiphoton sphere must be alternately.
 \item[(d)]The outermost circular light orbits are stable. Thus, the wormhole has at least one photon sphere on or off the throat.  
 \item[(e)]In a degenerated case, i.e., $V_\mathrm{eff}^{\prime \prime}(x_\mathrm{th},b)=0$, the theorems (a)-(d) hold if we read $V_\mathrm{eff}^{\prime \prime }(x_\mathrm{th},b)$ in the theorems as $V_\mathrm{eff}^{\prime \prime \prime}(x_\mathrm{th},b)$. In higher-order degenerated cases 
 $V_\mathrm{eff}^{\prime \prime}(x_\mathrm{th},b)=V_\mathrm{eff}^{\prime \prime \prime}(x_\mathrm{th},b)=V_\mathrm{eff}^{\prime \prime \prime \prime}(x_\mathrm{th},b)=\cdots =V_\mathrm{eff}^{(m)}(x_\mathrm{th},b)=0$, where $V_\mathrm{eff}^{(m)}(x,b)$ is the $m$th derivative of the effective potential, the theorems (a)-(d) hold if we read $V_\mathrm{eff}^{\prime \prime }(x_\mathrm{th},b)$ in the theorems as $V_\mathrm{eff}^{(m+1)}(x_\mathrm{th},b)$.
\end{description}
Note that more than one photon spheres and antiphoton spheres on and/or off the throat can form degenerated photon spheres or degenerated antiphoton spheres. The theorems (a)-(e) are true if we regard the degenerated photon spheres and the degenerated antiphoton spheres as the photon spheres and the antiphoton spheres before the degenerations.  
See Refs.~\cite{Chiba:2017nml,Tsukamoto:2020iez,Tsukamoto:2020uay,Tsukamoto:2020bjm,Tsukamoto:2023rzd,Zhang:2024sgs} for the details of the degenerated photon spheres in various spacetimes including wormhole spacetimes.

\section{Examples}
In this section, we check the theorems (a)-(e)  
in the Simpson-Visser black-bounce spacetime~\cite{Simpson:2018tsi},
the Damour-Solodukhin wormhole spacetime~\cite{Damour:2007ap}, 
and the Reissner-Nordstr\"{o}m black-hole-like wormhole spacetime~\cite{Lemos:2008cv,Tsukamoto:2019ihj}.

\subsection{Simpson-Visser black-bounce spacetime}
Simpson and Visser suggested a black-bounce spacetime~\cite{Simpson:2018tsi}
with metric functions 
\begin{eqnarray}
&&A(r)=1-\frac{2M}{\sqrt{x^2+a^2}}, \nonumber\\
&&r(x)=\sqrt{x^2+a^2}, 
\end{eqnarray}
where $M$ is a positive ADM mass, $a$ is a nonnegative constant, and the radial coordinate is defined in $-\infty <x< \infty$. 
It is a Schwarzshild spacetime for $a=0$, a regular black hole spacetime for $a<2M$, 
a one-way traversable wormhole spacetime for $a=2M$, 
and a two-way traversable wormhole for $a>2M$.
In the case of no ADM mass $M=0$, the metric becomes
\begin{eqnarray}
\mathrm{d}s^2=-\mathrm{d}t^2+ \mathrm{d}x^2 +(x^2+a^2)\left( \mathrm{d}\theta^2+ \sin^2 \theta \mathrm{d}\varphi^2 \right), 
\end{eqnarray}
which is corresponds to the metric of the Ellis-Bronnikov wormhole~\cite{Ellis:1973yv,Bronnikov:1973fh} with no ADM masses.    
In Ref.~\cite{Dey:2008kn}, 
Dey and Sen claimed that
the Ellis-Bronnikov wormhole with no ADM masses has no photon sphere on its throat
but it contradicts papers by several authors~\cite{Ellis:1973yv,Chetouani_Clement_1984,Perlick:2003vg,Perlick_2004_Living_Rev,Muller:2004dq,Nandi:2006ds,Muller:2008zza,Bhattacharya:2010zzb,Gibbons:2011rh,Nakajima:2012pu,Tsukamoto:2012xs,Ohgami:2015nra,Tsukamoto:2016zdu,Tsukamoto:2016qro,Tsukamoto:2017edq,Bugaev:2023mlc}.

The effective potential $V_\mathrm{eff}(x,b)$ of the ray in the Simpson-Visser spacetime is given by
\begin{eqnarray}
V_\mathrm{eff}(x,b)= E^2  \left[ \left(1-\frac{2M}{\sqrt{x^2+a^2}}\right) \frac{b^2}{x^2+a^2} -1 \right].
\end{eqnarray}
From the first to fourth derivatives of the effective potential are obtained as 
\begin{eqnarray}
V_\mathrm{eff}^{\prime}(x,b)=   \frac{2 E^2 b^2 x \left(3M-\sqrt{x^2+a^2}\right)}{ \left( x^2+a^2 \right)^\frac{5}{2}}, 
\end{eqnarray}
\begin{eqnarray}
V_\mathrm{eff}^{\prime \prime}(x,b)
&=&\frac{2 E^2 b^2}{ \left( x^2+a^2 \right)^\frac{7}{2}} 
\left[ -3x^2 \left(4M-\sqrt{x^2+a^2}\right) \right. \nonumber\\
&&\left. +a^2 \left(3M-\sqrt{x^2+a^2}\right) \right], 
\end{eqnarray}
\begin{eqnarray}
V_\mathrm{eff}^{\prime \prime \prime}(x,b)
&=&\frac{6 E^2 b^2 x}{ \left( x^2+a^2 \right)^\frac{9}{2}} 
\left[ 4x^2 \left(5M-\sqrt{x^2+a^2}\right) \right. \nonumber\\
&&\left. +a^2 \left(-15M+4\sqrt{x^2+a^2}\right) \right], 
\end{eqnarray}
\begin{eqnarray}
V_\mathrm{eff}^{\prime \prime \prime \prime}(x,b)
&=&\frac{6 E^2 b^2 }{ \left( x^2+a^2 \right)^\frac{11}{2}} 
\left[ 20 a^2 x^2 \left(9M-2\sqrt{x^2+a^2}\right) \right. \nonumber\\
&&+20 x^4 \left(-6M+\sqrt{x^2+a^2}\right) \nonumber\\
&&\left. +a^4 \left(-15M+4\sqrt{x^2+a^2}\right) \right].
\end{eqnarray}

For $0 \leq a< 3M$, we get 
\begin{eqnarray}
V_\mathrm{eff}(x_\mathrm{m},b_\mathrm{m})=V_\mathrm{eff}^{\prime}(x_\mathrm{m},b_\mathrm{m})=0
\end{eqnarray}
and 
\begin{eqnarray}
V_\mathrm{eff}^{\prime \prime}(x_\mathrm{m},b_\mathrm{m})=-\frac{2E^2\left( 9M^2-a^2 \right)}{27M^4}<0
\end{eqnarray}
where
$x_\mathrm{m}$ is the radius of the circular light orbit given by   
\begin{eqnarray}
x_\mathrm{m} \equiv \sqrt{9M^2-a^2}
\end{eqnarray}
and $b_\mathrm{m}$ is the critical impact parameter given by 
\begin{eqnarray}
b_\mathrm{m} \equiv \pm 3\sqrt{3}M.
\end{eqnarray}
Thus, the black hole spacetime has one photon sphere at $x=x_\mathrm{m}$ for $0 \leq a< 2M$ 
and the wormhole has two photon spheres at $x=\pm x_\mathrm{m}$ which are off the throat in the both regions of the throat for $2M \leq a< 3M$.   
For $2M < a< 3M$,  
there is an antiphoton sphere on the throat at $x=x_\mathrm{th}=0$ since 
we get 
\begin{eqnarray}
V_\mathrm{eff}(x_\mathrm{th},b_\mathrm{th})=V_\mathrm{eff}^{\prime}(x_\mathrm{th},b_\mathrm{th})=0
\end{eqnarray}
and 
\begin{eqnarray}
V_\mathrm{eff}^{\prime \prime }(x_\mathrm{th},b_\mathrm{th})= \frac{2E^2b_\mathrm{th}^2 \left(3M-a \right)}{a^5} >0,
\end{eqnarray}
where $b_\mathrm{th}$ is the critical impact parameter obtained as 
\begin{eqnarray}
b_\mathrm{th} = \pm \sqrt{ \frac{a^3}{a-2M}}.
\end{eqnarray}
For $a>3M$,  
the wormhole has a photon sphere on the throat at $x=x_\mathrm{th}=0$ because of 
\begin{eqnarray}
V_\mathrm{eff}(x_\mathrm{th},b_\mathrm{th})=V_\mathrm{eff}^{\prime}(x_\mathrm{th},b_\mathrm{th})=0
\end{eqnarray}
and 
\begin{eqnarray}
V_\mathrm{eff}^{\prime \prime }(x_\mathrm{th},b_\mathrm{th})= \frac{2E^2b_\mathrm{th}^2 \left(3M-a \right)}{a^5} <0.
\end{eqnarray}
For a degenerated case $a= 3M$,
from 
\begin{eqnarray}
V_\mathrm{eff}(x_\mathrm{th},b_\mathrm{th})
&=&V_\mathrm{eff}^{\prime}(x_\mathrm{th},b_\mathrm{th})=V_\mathrm{eff}^{\prime \prime }(x_\mathrm{th},b_\mathrm{th}) \nonumber\\
&=&V_\mathrm{eff}^{\prime \prime \prime}(x_\mathrm{th},b_\mathrm{th})=0
\end{eqnarray}
and 
\begin{eqnarray}
V_\mathrm{eff}^{\prime \prime \prime \prime}(x_\mathrm{th},b_\mathrm{th})=-\frac{6E^2b_\mathrm{th}^2}{a^6}<0,
\end{eqnarray}
the wormhole has a degenerated photon sphere on the throat. 

Therefore, we conclude that 
the wormhole spacetime has two photon spheres at $x=\pm x_\mathrm{m}=\pm \sqrt{9M^2-a^2}$ which are off the throat in the both regions of the throat and one antiphoton sphere on the throat at $x=x_\mathrm{th}=0$ for $2M \leq a<3M$,
it has one degenerated photon sphere on the throat at $x=x_\mathrm{th}=0$ for $a=3M$, 
and it has one photon sphere on the throat at $x_\mathrm{th}=0$ for $a>3M$.
For $0 \leq a<2M$, the black hole spacetime has one photon sphere at $x=x_\mathrm{m}=\sqrt{9M^2-a^2}$.
It is consistent with the theorems (a)-(e) and with results in Ref.~\cite{Tsukamoto:2020bjm}
while the antiphoton sphere and the photon sphere on the throat were overlooked in Refs.~\cite{Simpson:2018tsi,Nascimento:2020ime,Guerrero:2021ues,Bronnikov:2021liv,Combi:2024ehi}.  
We plot $x_\mathrm{m}/M$ and $x_\mathrm{th}/M$ and their stability in Fig.~1. 
\begin{figure}[htbp]
\begin{center}
\includegraphics[width=80mm]{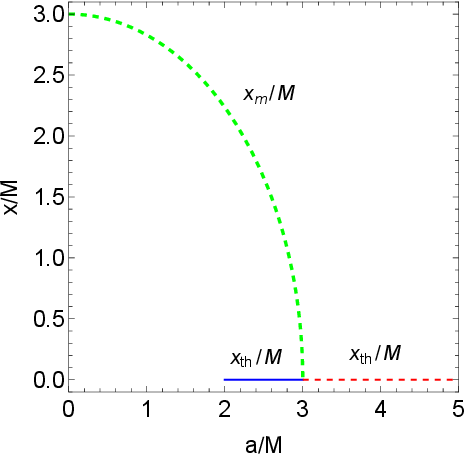}
\end{center}
\caption{
The specific radii of two photon spheres off the throat $x_\mathrm{m}/M$ for $2M \leq a < 3M$ denoted by thick-dashed~(green) curve, 
and the specific radii of a photon sphere for $2M \leq a < 3M$ and an antiphoton sphere for $a \geq 3M$ on the throat $x_\mathrm{th}/M$ denoted by thin-dashed~(red) and thin-solid~(blue) curves, respectively, in the Simpson-Visser wormhole spacetime are shown.
Dashed and solid curves denote photon spheres and antiphoton spheres, respectively.   
Note that the thick-dashed~(green) curve for $a<2M$ denotes the specific radius of a photon sphere $x_\mathrm{m}/M$ in not the two-way wormhole spacetime but the black hole spacetime or the one-way wormhole spacetime.}
\end{figure}

\subsection{Damour-Solodukhin wormhole spacetime}
A Damour-Solodukhin wormhole~\cite{Damour:2007ap} was suggested as a black hole mimicker 
with a metric 
\begin{eqnarray}
\mathrm{d}s^2
&=&- \left( 1-\frac{2M}{r}+\lambda^2 \right) \mathrm{d}\tilde{t}^2 \nonumber\\
&&+\frac{\mathrm{d}r^2}{1-\frac{2M}{r}} +r^2 \left( \mathrm{d}\theta^2+ \sin^2 \theta \mathrm{d}\varphi^2 \right), 
\end{eqnarray}
where $M>0$ and $\lambda > 0$ are assumed.
A tidal force~\cite{Lemos:2008cv}, a particle collision~\cite{Tsukamoto:2019ihj},
gravitational lensing~\cite{Nandi:2018mzm,Ovgun:2018fnk,Bhattacharya:2018leh,Ovgun:2018swe,Ovgun:2018oxk,Tsukamoto:2020uay,Tsukamoto:2023rzd}, 
quasinormal modes and gravitational waves~\cite{Bueno:2017hyj,Volkel:2018hwb},
the emissions~\cite{Karimov:2019qco} and images of accretion disks~\cite{Paul:2019trt}, and
a shadow~\cite{Amir:2018pcu,Tsukamoto:2020uay,Tsukamoto:2023rzd} in the Damour-Solodukhin wormhole spacetime were studied.

The wormhole mass is determined not only the parameter $M$ but it is determined by the parameters $M$ and $\lambda$~\cite{Tsukamoto:2023rzd}
while the contribution of $\lambda$ to the mass is often overlooked~\cite{Vagnozzi:2022moj}.
By transforming the time coordinate $\tilde{t}$ into $t\equiv \sqrt{1+\lambda^2} \tilde{t}$,
we obtain the metric in the coordinates $(t, r, \theta, \varphi)$ as  
\begin{eqnarray}
\mathrm{d}s^2
&=&-\left[ 1-\frac{2M}{(1+\lambda^2)r} \right] \mathrm{d}t^2  \nonumber\\
&&+\frac{\mathrm{d}r^2}{1-\frac{2M}{r}} +r^2 \left( \mathrm{d}\theta^2+ \sin^2 \theta \mathrm{d}\varphi^2 \right)
\end{eqnarray}
and we can see the contribution of $\lambda$ to the mass.
The spacetime is asymptotically flat and it has a throat at $r= r_{\mathrm{th}} \equiv 2M$ but the radial coordinate $r$ does not cover the other side of the throat.
We set $\theta=\pi/2$ without loss of generality.

We check the photon and antiphoton spheres off the throat. 
The trajectory of a ray can be expressed by
\begin{eqnarray}\label{eq:trajectory9}
\dot{r}^2+v(r,b)=0,
\end{eqnarray}
where the effective potential $v(r)$ is given by
\begin{eqnarray}
v(r,b)\equiv  \frac{E^2}{1-\frac{2M}{r}} \left( \frac{b^2}{r^2} - \frac{1}{1-\frac{2M}{(1+\lambda^2)r}} \right).
\end{eqnarray}
We obtain 
\begin{eqnarray}
v(r_\mathrm{m},b_\mathrm{m})=\frac{\mathrm{d}v(r_\mathrm{m},b_\mathrm{m})}{\mathrm{d}r}=0,
\end{eqnarray}
and 
\begin{eqnarray}
\frac{\mathrm{d}^2v(r_\mathrm{m},b_\mathrm{m})}{\mathrm{d}r^2}<0
\end{eqnarray}
where
$r_\mathrm{m}$ is the radii of two photon spheres off the throat in two regions against the throat given by   
\begin{eqnarray}
r_\mathrm{m} \equiv \frac{3M}{1+\lambda^2}
\end{eqnarray}
and $b_\mathrm{m}$ is the critical impact parameters given by 
\begin{eqnarray}
b_\mathrm{m} \equiv \pm \sqrt{3}r_\mathrm{m}
\end{eqnarray}
in straight-forward calculations. 
The wormhole has two photon spheres at $r=r_\mathrm{m}$ which are off the throat 
if a condition $r_\mathrm{m} > r_\mathrm{th}$ or $\lambda < \lambda_\mathrm{th} \equiv 1/\sqrt{2}$ is satisfied 
and it has no photon sphere off the throat if $r_\mathrm{m} < r_\mathrm{th}$ or $\lambda > \lambda_\mathrm{th}$ holds.  

We investigate the photon sphere and the antiphoton sphere on the throat
by using a proper length $\rho$ from the throat defined by
\begin{eqnarray}
\left| \rho \right| 
&\equiv& 
\int^r_{r_{\mathrm{th}}} \sqrt{ 1-\frac{2M}{r} } \mathrm{d}r  \nonumber\\
&=&\sqrt{r^2-2 M r } \nonumber\\
&&+M \log \left( \frac{r-M+\sqrt{r^2-2 M r }}{M} \right) \nonumber\\
\end{eqnarray}
because the radial coordinate $x$ is slightly complicated in the Damour-Solodukhin wormhole spacetime.
By using the coordinates $(t, \rho, \theta, \varphi)$, where the domain of $\rho$ is $-\infty<\rho<\infty$, the metric is written in
\begin{eqnarray}
\mathrm{d}s^2
&=&- \left[ 1-\frac{2M}{(1+\lambda^2)r(\rho)} \right] \mathrm{d}t^2 \nonumber\\
&&+\mathrm{d}\rho^2 +r^2(\rho) \left( \mathrm{d}\theta^2+ \sin^2 \theta \mathrm{d}\varphi^2 \right)
\end{eqnarray}
and the throat is at $\rho=\rho_\mathrm{th}\equiv 0$.
The trajectory of the ray can be rewritten as
\begin{eqnarray}\label{eq:trajectory3}
\dot{\rho}^2+V(\rho,b)=0,
\end{eqnarray}
where the effective potential $V(\rho,b)$ is defined by
\begin{eqnarray}
V(\rho,b)\equiv E^2 \left( \frac{b^2}{r^2(\rho)} - \frac{1}{1-\frac{2M}{(1+\lambda^2)r(\rho)} } \right).
\end{eqnarray}
The ray with the critical impact parameter 
\begin{eqnarray}
b=b_\mathrm{th} \equiv \pm \sqrt{ \frac{r_\mathrm{th}^2}{1-\frac{2M}{(1+\lambda^2)r_\mathrm{th}}}} 
= \pm \frac{2M \sqrt{1+\lambda^2}}{\lambda}
\end{eqnarray}
satisfies 
\begin{eqnarray}
V(\rho_\mathrm{th}, b_\mathrm{th})=\frac{\mathrm{d}V(\rho_\mathrm{th}, b_\mathrm{th})}{\mathrm{d}\rho}=0
\end{eqnarray}
on the throat.
We obtain the higher-order derivatives of the effective potential on the throat as 
\begin{eqnarray}
\frac{\mathrm{d^2}V(\rho_\mathrm{th}, b_\mathrm{th})}{\mathrm{d}\rho^2}= \frac{E^2 \left(1-2\lambda^2 \right) \left(1+\lambda^2 \right)}{8M^2\lambda^4} >0
\end{eqnarray}
for $\lambda < \lambda_\mathrm{m}\equiv 1/\sqrt{2}$,  
\begin{eqnarray}
\frac{\mathrm{d^2}V(\rho_\mathrm{th}, b_\mathrm{th})}{\mathrm{d}\rho^2} <0
\end{eqnarray}
for $\lambda > \lambda_\mathrm{m}$, 
and
\begin{eqnarray}
&&\frac{\mathrm{d^2}V(\rho_\mathrm{th}, b_\mathrm{th})}{\mathrm{d}\rho^2} = \frac{\mathrm{d^3}V(\rho_\mathrm{th}, b_\mathrm{th})}{\mathrm{d}\rho^3}=0, \\
&&\frac{\mathrm{d^4}V(\rho_\mathrm{th}, b_\mathrm{th})}{\mathrm{d}\rho^4}= -\frac{27E^2}{32M^2} <0
\end{eqnarray}
for the degenerated case $\lambda = \lambda_\mathrm{m}$.
Thus, the wormhole has the antiphoton sphere for $\lambda < \lambda_\mathrm{m}$,
the photon sphere for $\lambda > \lambda_\mathrm{m}$,
and the degenerated photon sphere for $\lambda = \lambda_\mathrm{m}$ on the throat.
We plot $r_\mathrm{m}/M$ and $r_\mathrm{th}/M$ and their stability in Fig.~2. 
\begin{figure}[htbp]
\begin{center}
\includegraphics[width=80mm]{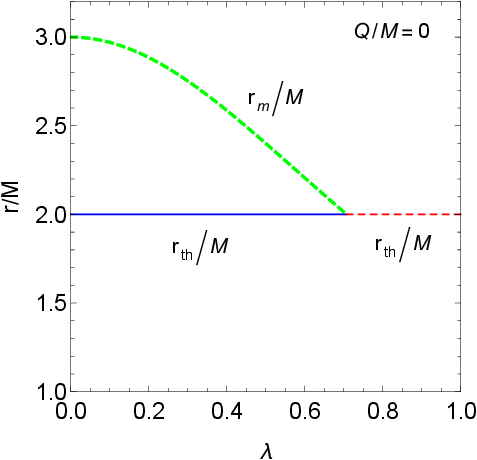}
\end{center}
\caption{
The specific radii of two photon spheres off the throat $r_\mathrm{m}/M$ denoted by thick-dashed~(green) curve, 
and the throat $r_\mathrm{th}/M$ denoted by thin-dashed (red) and thin-solid~(blue) curves in the Damour-Solodukhin wormhole spacetime are shown.
Dashed and solid curves denote photon spheres and antiphoton spheres, respectively.   
We note that it is a vanishing charged case $Q/M=0$, where $Q$ defined in subsection~III-C. 
}
\end{figure}

We summarize the photon spheres and the antiphoton spheres on an off the throat in the Damour-Solodukhin wormhole spacetime.
It has two photon spheres off the throat and one antiphoton sphere on the throat for $\lambda < 1/\sqrt{2}$,
it has one degenerated photon sphere which made from two photon sphere and one antiphoton sphere on the throat for $\lambda = 1/\sqrt{2}$,
and it has one photon sphere on the throat $\lambda > 1/\sqrt{2}$.
This result is consistent with the theorems (a)-(e) and Refs.~\cite{Tsukamoto:2020uay,Tsukamoto:2023rzd}.  
We note that the (anti)photon sphere on the throat is often neglected~\cite{Nandi:2018mzm,Ovgun:2018fnk,Bhattacharya:2018leh,Ovgun:2018swe,Ovgun:2018oxk,Vagnozzi:2022moj}.

\subsection{Reissner-Nordstr\"{o}m black-hole-like wormhole}
We confirm that the photon spheres and the antiphoton spheres 
in a Reissner-Nordstr\"{o}m black-hole-like wormhole spacetime
or a charged Damour-Solodukhin wormhole spacetime~\cite{Lemos:2008cv,Tsukamoto:2019ihj}
with a metric, 
in coordinates $(\tilde{t}, r, \theta, \varphi)$,
\begin{eqnarray}
\mathrm{d}s^2
&=&- \left( 1-\frac{2M}{r}+\frac{Q^2}{r^2}+\lambda^2 \right) \mathrm{d}\tilde{t}^2 \nonumber\\
&&+\frac{\mathrm{d}r^2}{1-\frac{2M}{r}+\frac{Q^2}{r^2}} +r^2 \left( \mathrm{d}\theta^2+ \sin^2 \theta \mathrm{d}\varphi^2 \right), \nonumber\\
\end{eqnarray}
where we assume $M>0$, $\lambda > 0$, and $\left| Q \right| \leq M$.
A tidal force~\cite{Lemos:2008cv} and a particle collision~\cite{Tsukamoto:2019ihj} in the Reissner-Nordstr\"{o}m black-hole-like wormhole spacetime was investigated.
By transforming the time coordinate $\tilde{t}$ into $t\equiv \sqrt{1+\lambda^2} \tilde{t}$, we obtain the metric in the coordinates $(t, r, \theta, \varphi)$ as  
\begin{eqnarray}
\mathrm{d}s^2=-A(r) \mathrm{d}t^2+\frac{\mathrm{d}r^2}{B(r)} +r^2 \left( \mathrm{d}\theta^2+ \sin^2 \theta \mathrm{d}\varphi^2 \right), 
\end{eqnarray}
where $A(r)$ and $B(r)$ are given by
\begin{eqnarray}
A(r)= 1-\frac{2M}{(1+\lambda^2)r}+\frac{Q^2}{(1+\lambda^2)r^2} 
\end{eqnarray}
and 
\begin{eqnarray}
B(r)= 1-\frac{2M}{r}+\frac{Q^2}{r^2},
\end{eqnarray}
respectively.
We note that the spacetime is asymptotically flat.
A throat is at $r= r_{\mathrm{th}} \equiv M+\sqrt{M^2-Q^2}$ but the radial coordinate $r$ does not cover the other side of the throat.
We set $\theta=\pi/2$ without loss of generality.

The trajectory of a ray can be expressed by
\begin{eqnarray}\label{eq:trajectory9}
\dot{r}^2+v(r,b)=0,
\end{eqnarray}
where the effective potential $v(r)$ is given by
\begin{eqnarray}
v(r,b)\equiv E^2 B(r) \left( \frac{b^2}{r^2} - \frac{1}{A(r)} \right).
\end{eqnarray}
There are two photon (antiphoton) spheres at $r=r_\mathrm{m}$ ($r=r_\mathrm{a}$)
in both regions against the throat, 
if 
$\lambda < \lambda_\mathrm{ma}$ and $r_\mathrm{m} > r_\mathrm{th}$ ($\lambda < \lambda_\mathrm{ma}$ and $r_\mathrm{a} > r_\mathrm{th}$) are satisfied.
Here, $r_\mathrm{m}$ and $r_\mathrm{a}$ are given by   
\begin{eqnarray}
r_\mathrm{m} \equiv \frac{3M + \sqrt{9 M^2-8 \left( 1+\lambda^2 \right) Q^2}}{2 \left(1+\lambda^2 \right)}
\end{eqnarray}
and 
\begin{eqnarray}
r_\mathrm{a} \equiv \frac{3M - \sqrt{9 M^2-8 \left( 1+\lambda^2 \right) Q^2}}{2 \left(1+\lambda^2 \right)},
\end{eqnarray}
respectively, and 
$\lambda_\mathrm{ma}$ is defined as  
\begin{eqnarray}
\lambda_\mathrm{ma} \equiv \frac{\sqrt{9M^2-8Q^2}}{2\sqrt{2} \left| Q \right|}.
\end{eqnarray}
We find that $r_\mathrm{m} = r_\mathrm{th}$ or $r_\mathrm{a} = r_\mathrm{th}$ holds if $\lambda=\lambda_\mathrm{th}$, 
where $\lambda_\mathrm{th}$ is defined by
\begin{eqnarray}
\lambda_\mathrm{th} \equiv \sqrt{1+ \frac{M \left(\sqrt{M^2-Q^2}-M\right)}{Q^2}}
\end{eqnarray} 
for $0< \left| Q \right| \leq M$ 
and $\lambda_\mathrm{th} \equiv 1/\sqrt{2}$ for $Q=0$.
Figure~3 shows $\lambda_\mathrm{th}$ and $\lambda_\mathrm{ma}$ as the functions of $\left|Q\right|/M$.
We find that $\lambda=1/\sqrt{5} \sim 0.447$ and $\left| Q \right|/M=\sqrt{15}/4 \sim 0.968$ is the only solution of $\lambda=\lambda_\mathrm{th}=\lambda_\mathrm{ma}$.
\begin{figure}[htbp]
\begin{center}
\includegraphics[width=80mm]{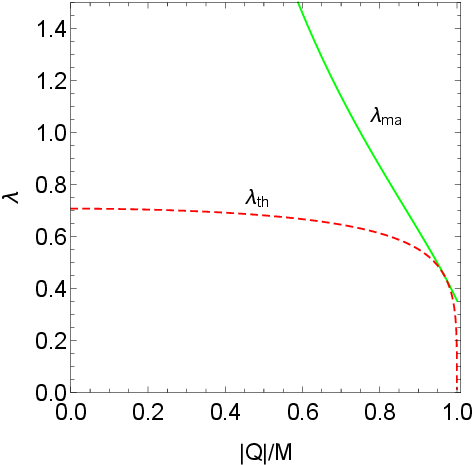}
\end{center}
\caption{
$\lambda_\mathrm{ma}$ and $\lambda_\mathrm{th}$ as the functions of $\left| Q \right|/M$ are shown by
as solid~(green) and dashed~(red) curves, respectively.
Given $\left| Q \right|/M$,
We find that $\lambda_\mathrm{ma}$ is greater than or equal to $\lambda_\mathrm{th}$. We get $\lambda_\mathrm{th}=\lambda_\mathrm{ma}$ if and only if $\left| Q \right|/M=\sqrt{15}/4 \sim 0.968$.
}
\end{figure}
We can confirm 
\begin{eqnarray}
&&v(r_\mathrm{m},b_\mathrm{m})=\frac{\mathrm{d}v(r_\mathrm{m},b_\mathrm{m})}{\mathrm{d}r}=0,\\
&&v(r_\mathrm{a},b_\mathrm{a})=\frac{\mathrm{d}v(r_\mathrm{a},b_\mathrm{a})}{\mathrm{d}r}=0
\end{eqnarray}
and 
\begin{eqnarray}
&&\frac{\mathrm{d}^2v(r_\mathrm{m},b_\mathrm{m})}{\mathrm{d}r^2}<0,\\
&&\frac{\mathrm{d}^2v(r_\mathrm{a},b_\mathrm{a})}{\mathrm{d}r^2}>0,
\end{eqnarray}
where $b_\mathrm{m}$ and $b_\mathrm{a}$ are the critical impact parameters given by 
\begin{eqnarray}
&&b_\mathrm{m} \equiv \pm \sqrt{\frac{r_\mathrm{m}^2}{A(r_\mathrm{m})}},\\
&&b_\mathrm{a} \equiv \pm \sqrt{\frac{r_\mathrm{a}^2}{A(r_\mathrm{a})}},
\end{eqnarray}
in straight-forward calculations. 
There are two degenerated photon spheres at $r=r_\mathrm{m}=r_\mathrm{a}$ off the throat in both regions against the throat,
in a degenerated case with $\lambda=\lambda_\mathrm{ma}$ and $r_\mathrm{m}=r_\mathrm{a} > r_\mathrm{th}$, we obtain
\begin{eqnarray}
&&v(r_\mathrm{m},b_\mathrm{m})=v(r_\mathrm{a},b_\mathrm{a})=\frac{\mathrm{d}v(r_\mathrm{m},b_\mathrm{m})}{\mathrm{d}r}
=\frac{\mathrm{d}v(r_\mathrm{a},b_\mathrm{a})}{\mathrm{d}r} \nonumber\\
&&=\frac{\mathrm{d}^2v(r_\mathrm{m},b_\mathrm{m})}{\mathrm{d}r^2} =\frac{\mathrm{d}^2v(r_\mathrm{a},b_\mathrm{a})}{\mathrm{d}r^2}=0
\end{eqnarray}
and
\begin{eqnarray}
&&\frac{\mathrm{d}^3v(r_\mathrm{m},b_\mathrm{m})}{\mathrm{d}r^3} 
=\frac{\mathrm{d}^3v(r_\mathrm{a},b_\mathrm{a})}{\mathrm{d}r^3} \nonumber\\
&&=\frac{243 E^2 \left(15 M^5-16 M^3 Q^2\right)}{64 Q^8}<0.
\end{eqnarray}
We plot $r_\mathrm{m}/M$, $r_\mathrm{a}/M$, $r_\mathrm{th}/M$ in Fig.~4 and we show examples of the second derivatives of the effective potentials in Fig.~5. 
\begin{figure}[htbp]
\begin{center}
\includegraphics[width=64mm]{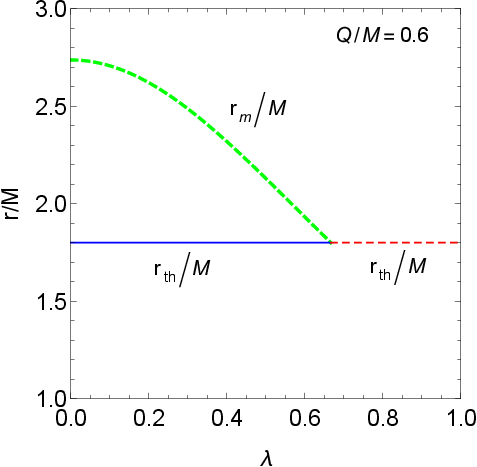}
\includegraphics[width=64mm]{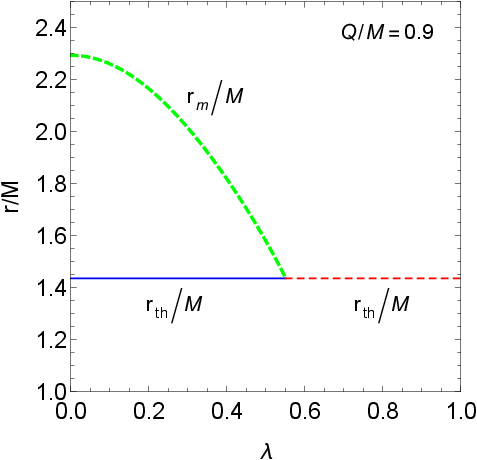}
\includegraphics[width=64mm]{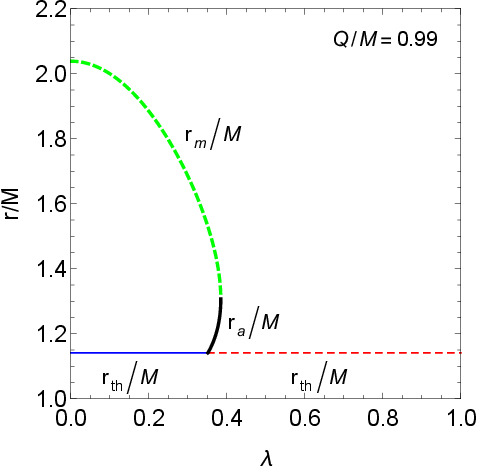}
\end{center}
\caption{
The specific radii of two photon spheres off the throat $r_\mathrm{m}/M$, 
two antiphoton spheres off the throat $r_\mathrm{a}/M$, 
and a photon sphere or an antiphoton sphere on the throat $r_\mathrm{th}/M$ 
when $Q/M=0.6, 0.9$, and $0.99$ are shown in top, middle, and bottom panels, respectively. 
Thick-dashed~(green) and thick-solid~(black) curves denote 
the specific radii of two photon spheres off the throat $r_\mathrm{m}/M$ and 
two antiphoton spheres off the throat $r_\mathrm{a}/M$, respectively. 
Thin-dashed (red) and thin-solid~(blue) curves denote the specific radius of a photon sphere 
and an antiphoton sphere on the throat $r_\mathrm{th}/M$, respectively.
Dashed and solid curves denote photon spheres and antiphoton spheres, respectively.   
}
\end{figure}
\begin{figure}[htbp]
\begin{center}
\includegraphics[width=80mm]{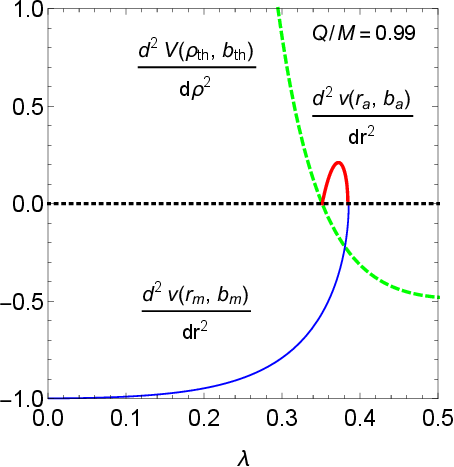}
\end{center}
\caption{
The second derivatives of the effective potentials for $Q/M=0.99$ are shown.
Thin-solid~(blue) and thick-solid~(red) curves denote 
the second derivatives of effective potentials for the photon spheres and the antiphoton spheres off the throat, 
respectively.
A thick-dashed~(green) curve denotes the one for another photon sphere or another antiphoton sphere on the throat.
We confirm that the curves have two zero points at $\lambda=\lambda_\mathrm{th}$ and $\lambda=\lambda_\mathrm{ma}$.  
}
\end{figure}

From the above, we obtain the photon spheres and the antiphotons spheres off the throat in the both regions against the throat in the Reissner-Nordstr\"{o}m black-hole-like wormhole. 
\begin{description} 
  \item[(1)] For $\left| Q \right|/M \leq \sqrt{15}/4$ and $\lambda<\lambda_\mathrm{th}$, the wormhole has two photon spheres off the throat.
  \item[(2)] For $\left| Q \right|/M \leq \sqrt{15}/4$ and $\lambda \geq \lambda_\mathrm{th}$, it has neither photon sphere nor antiphoton sphere off the throat.
  \item[(3)] For $\sqrt{15}/4 < \left| Q \right|/M \leq 1$ and $\lambda < \lambda_\mathrm{th}$, it has two photon spheres off the throat.
  \item[(4)] For $\sqrt{15}/4 < \left| Q \right|/M \leq 1$ and $\lambda = \lambda_\mathrm{th}$, it has neither photon sphere nor antiphoton sphere off the throat.
  \item[(5)] For $\sqrt{15}/4 < \left| Q \right|/M \leq 1$ and $\lambda_\mathrm{th}<\lambda<\lambda_\mathrm{ma}$, it has two photon spheres and two antiphoton spheres off the throat.
  \item[(6)] For $\sqrt{15}/4 < \left| Q \right|/M \leq 1$ and $\lambda=\lambda_\mathrm{ma}$, it has two degenerated photon spheres which are made from two photon spheres and two antiphoton spheres off the throat.
  \item[(7)] For $\sqrt{15}/4 < \left| Q \right|/M \leq 1$ and $\lambda>\lambda_\mathrm{ma}$, it has neither photon sphere nor antiphoton sphere off the throat.
\end{description} 

We use rather a proper length $\rho$ from the throat as new radial coordinate to cover over the Reissner-Nordstr\"{o}m black-hole-like wormhole spacetime 
than the radial coordinate $x$ since the latter is really complicated. 
We obtain the proper length as
\begin{eqnarray}
\left| \rho \right| 
&\equiv& 
\int^r_{r_{\mathrm{th}}} \frac{\mathrm{d}r}{\sqrt{B(r)}} \nonumber\\
&=&\sqrt{r^2-2 M r+Q^2} \nonumber\\
&&+M \log \left( \frac{r-M+\sqrt{r^2-2 M r+Q^2}}{\sqrt{M^2-Q^2}} \right). \nonumber\\
\end{eqnarray}
By using the coordinates $(t, \rho, \theta, \varphi)$, the metric is written in
\begin{eqnarray}
\mathrm{d}s^2
&=&- \left[ 1-\frac{2M}{(1+\lambda^2)r(\rho)}+\frac{Q^2}{(1+\lambda^2)r^2(\rho)} \right] \mathrm{d}t^2 \nonumber\\
&&+\mathrm{d}\rho^2 +r^2(\rho) \left( \mathrm{d}\theta^2+ \sin^2 \theta \mathrm{d}\varphi^2 \right). \nonumber\\
\end{eqnarray}
Note that the throat is at $\rho=\rho_\mathrm{th}\equiv 0$ and we can extend the domain of the radial coordinate $\rho$ into $-\infty<\rho<\infty$ because of the $Z_2$ symmetry of the wormhole spacetime.
The trajectory of the ray can be rewritten as
\begin{eqnarray}\label{eq:trajectory3}
\dot{\rho}^2+V(\rho,b)=0,
\end{eqnarray}
where the effective potential $V(\rho,b)$ is defined by
\begin{eqnarray}
V(\rho,b)\equiv E^2 \left( \frac{b^2}{r^2(\rho)} - \frac{1}{A(r(\rho))} \right).
\end{eqnarray}
The ray with the critical impact parameter 
\begin{eqnarray}
b=b_\mathrm{th} \equiv \pm \sqrt{\frac{r_\mathrm{th}^2}{A(r_\mathrm{th})}}
\end{eqnarray}
satisfies 
\begin{eqnarray}
V(\rho_\mathrm{th}, b_\mathrm{th})=\frac{\mathrm{d}V(\rho_\mathrm{th}, b_\mathrm{th})}{\mathrm{d}\rho}=0.
\end{eqnarray}
The second derivative of the effective potential is obtained as 
\begin{eqnarray}
&&\frac{\mathrm{d^2}V(\rho_\mathrm{th}, b_\mathrm{th})}{\mathrm{d}\rho^2} \nonumber\\
&&=E^2 \left( -\frac{b_\mathrm{th}^2}{r_\mathrm{th}^3}+ \frac{1}{2A^2(r_\mathrm{th})} \frac{\mathrm{d}A(r_\mathrm{th})}{\mathrm{d}r}  \right) \frac{\mathrm{d}B(r_\mathrm{th})}{\mathrm{d}r} \nonumber\\
&&=\frac{-2 \left(\lambda ^2+1\right) \left(M \sqrt{M^2-Q^2}+M^2-Q^2\right)E^2 Z }{\lambda ^4 \left(\sqrt{M^2-Q^2}+M\right)^4 } \nonumber\\
&&\times \frac{1}{\left(2 M \sqrt{M^2-Q^2}+2 M^2-Q^2\right)^2},
\end{eqnarray}
where $Z$ is given by
\begin{eqnarray}
Z&\equiv& 5 M^2 Q^2+\left(3 M Q^2-4 M^3\right) \sqrt{M^2-Q^2}-4 M^4-Q^4 \nonumber\\
&&+\lambda ^2 \left[8 M^4+Q^4-8 M^2 Q^2 \right. \nonumber\\
&&\left.   +\left(8 M^3-4 M Q^2\right) \sqrt{M^2-Q^2}      \right]. 
\end{eqnarray}
In straight-forward calculations, we can confirm 
\begin{eqnarray}
\frac{\mathrm{d^2}V(\rho_\mathrm{th}, b_\mathrm{th})}{\mathrm{d}\rho^2}>0
\end{eqnarray} 
for $\lambda<\lambda_\mathrm{th}$,  
\begin{eqnarray}
\frac{\mathrm{d^2}V(\rho_\mathrm{th}, b_\mathrm{th})}{\mathrm{d}\rho^2}<0
\end{eqnarray} 
for $\lambda>\lambda_\mathrm{th}$, and 
\begin{eqnarray}
&&\frac{\mathrm{d^2}V(\rho_\mathrm{th}, b_\mathrm{th})}{\mathrm{d}\rho^2}=\frac{\mathrm{d^3}V(\rho_\mathrm{th}, b_\mathrm{th})}{\mathrm{d}\rho^3}=0, \\
&&\frac{\mathrm{d^4}V(\rho_\mathrm{th}, b_\mathrm{th})}{\mathrm{d}\rho^4}<0
\end{eqnarray}
for the degenerated case with $\lambda=\lambda_\mathrm{th}$ and $Q/M \neq \sqrt{15}/4$,
and 
\begin{eqnarray}
&&\frac{\mathrm{d^2}V(\rho_\mathrm{th}, b_\mathrm{th})}{\mathrm{d}\rho^2}=\frac{\mathrm{d^3}V(\rho_\mathrm{th}, b_\mathrm{th})}{\mathrm{d}\rho^3} \nonumber\\
&&=\frac{\mathrm{d^4}V(\rho_\mathrm{th}, b_\mathrm{th})}{\mathrm{d}\rho^4}=\frac{\mathrm{d^5}V(\rho_\mathrm{th}, b_\mathrm{th})}{\mathrm{d}\rho^5}=0, \\
&&\frac{\mathrm{d^6}V(\rho_\mathrm{th}, b_\mathrm{th})}{\mathrm{d}\rho^6}<0
\end{eqnarray}
for the degenerated case with $\lambda=\lambda_\mathrm{th}=\lambda_\mathrm{ma}=1/\sqrt{5}$ and $Q/M = \sqrt{15}/4$.
Thus, there is an antiphoton sphere on the throat for $\lambda<\lambda_\mathrm{th}$, 
a photon sphere on the throat for $\lambda>\lambda_\mathrm{th}$, and
a degenerated photon sphere on the throat for $\lambda=\lambda_\mathrm{th}$. 

We summarize our results on the Reissner-Nordstr\"{o}m black-hole-like wormhole. 
\begin{description} 
  \item[(i)] For $\left| Q \right|/M \leq \sqrt{15}/4$ and $\lambda<\lambda_\mathrm{th}$, the wormhole has two photon spheres off the throat and one antiphoton sphere on the throat.
  \item[(ii)] For $\left| Q \right|/M < \sqrt{15}/4$ and $\lambda = \lambda_\mathrm{th}$, it has one degenerated photon sphere which are made from two photon spheres and it has one antiphoton sphere on the throat.
  \item[(iii)] For $\left| Q \right|/M \leq \sqrt{15}/4$ and $\lambda > \lambda_\mathrm{th}$, it has one photon sphere on the throat.
  \item[(iv)] For $\left| Q \right|/M = \sqrt{15}/4$ and $\lambda = \lambda_\mathrm{th}=\lambda_\mathrm{ma}$, it has one degenerated photon sphere which are made from three photon spheres and it has two antiphoton sphere on the throat.
  \item[(v)] For $\sqrt{15}/4 < \left| Q \right|/M \leq 1$ and $\lambda < \lambda_\mathrm{th}$, it has two photon spheres off the throat and one antiphoton sphere on the throat.
  \item[(vi)] For $\sqrt{15}/4 < \left| Q \right|/M \leq 1$ and $\lambda = \lambda_\mathrm{th}$, it has one degenerated photon sphere which are made from two photon spheres and it has one antiphoton sphere on the throat.
  \item[(vii)] For $\sqrt{15}/4 < \left| Q \right|/M \leq 1$ and $\lambda_\mathrm{th}<\lambda<\lambda_\mathrm{ma}$, it has two photon spheres and two antiphoton spheres off the throat and one photon sphere on the throat.
  \item[(viii)] For $\sqrt{15}/4 < \left| Q \right|/M \leq 1$ and $\lambda=\lambda_\mathrm{ma}$, it has two degenerated photon spheres which are made from two photon spheres and two antiphoton spheres off the throat and it has one antiphoton sphere on the throat.
  \item[(ix)] For $\sqrt{15}/4 < \left| Q \right|/M \leq 1$ and $\lambda>\lambda_\mathrm{ma}$, it has one photon sphere on the throat.
\end{description} 
We have confirmed that this result is consistent with the theorems (a)-(e).

\section{Summary and Discussion}
A general asymptotically-flat, static, and spherical symmetrical wormhole without a thin shell has at least 
one unstable circular light orbit at some radius as pointed out by Perlick \cite{Perlick_2004_Living_Rev}. 
If we assume $Z_2$ symmetry against a throat, 
a stable or unstable circular light orbit is on the throat~\cite{Shaikh:2018oul,Bronnikov:2018nub} but it is often overlooked. 

We have revisited the circular light orbit of $Z_2$ symmetrical wormhole to fill following lacking points in Refs.~\cite{Perlick_2004_Living_Rev,Shaikh:2018oul,Bronnikov:2018nub}. 
In Ref.~\cite{Perlick_2004_Living_Rev}, the existence of unstable circular light orbit somewhere of any asymptotically-flat Morris-Thorne wormhole without a thin shell was pointed out but the fact that $Z_2$-symmetrical wormholes have circular light orbits on the throat was not pointed out. In Refs.~\cite{Shaikh:2018oul,Bronnikov:2018nub}, the light circular orbit on the throat of the $Z_2$-symmetrical wormhole was found. However, the authors of Ref.~\cite{Shaikh:2018oul} did not comment that asymptotically-flat, static, and spherical symmetrical, and $Z_2$ symmetrical wormholes without the thin shell always have the unstable light circular orbits on or off the throat. And the authors of Ref.~\cite{Bronnikov:2018nub} seem to overlook a case of the $Z_2$-symmetrical wormhole with stable light circular orbits on the throat. 
They were just small lacking points before the ring images or the shadows of the centers of the M87 and the Milky way have been reported by the Event Horizon Telescope Collaborations.  
Recently, however, they have been important problems to be revisited since
the ring images or the shadows of the centers of the M87 and the Milky way imply that they have the photon spheres.

In this work, we summarize the results as theorems (a)-(e).
We have categorized the numbers of the circular light orbits of the $Z_2$-symmetrical wormhole and their stability from the derivatives of an effective potential at the throat.
We note that the $Z_2$-symmetrical wormhole has at least one unstable circular light orbit on or off the throat
and that our results do not depend on gravitational theories and energy conditions.
We have given three examples of Simpson-Visser black-bounce spacetime, a Damour-Solodukhin wormhole, a Reissner-Nordstr\"{o}m black-hole-like wormhole or a charged Damour-Solodukhin wormhole to confirm our theorems (a)-(e).
In the theorems (a)-(e) and the examples, we have given complete treatments including the degenerated photon spheres made from the photon spheres and the antiphoton spheres on and off the throat.    

In this work, we have concentrated on static and spherical symmetric wormholes
but circular light orbits around axisymmetric wormholes are also a fascinated topic.
Maeda found that a class of the axisymmetric wormholes with the $Z_2$ symmetry has the circular light orbits on a throat~\cite{Maeda:2021wnl}.
The effects of a throat on circular light orbits 
or the shadow of rotating wormholes also escape notice sometimes~\cite{Nedkova:2013msa,Abdujabbarov:2016efm,Amir:2018pcu}
while the throat can change the shape of the shadow of the rotating wormholes~\cite{Shaikh:2018kfv,Kasuya:2021cpk}.
The extension of this work to a general axisymmetric wormhole is left as future work.

\section*{Acknowledgements}
The author is deeply grateful to Volker Perlick for his valuable comments on the existence theorem of a photon sphere of the general asymptotically-flat, 
static, and spherical symmetrical wormhole without the $Z_2$ symmetry 
and to Hideki Maeda for his comments on the circular light orbits of axisymmetric wormhole and $Z_2$-symmetrical wormhole.    
The author also is grateful to an anonymous referee for his or her valuable comments.

\end{document}